\documentclass[12pt]{article}
\usepackage{xr-hyper}
\usepackage{changepage}

    \usepackage[dvipsnames]{xcolor}
    \usepackage[T1]{fontenc}
	\usepackage[utf8]{inputenc}
	\usepackage[english]{babel}
	\usepackage{amsfonts}
	\usepackage{amssymb}
	    \usepackage{amsmath}
	\usepackage{adjustbox}
	\usepackage{threeparttable}
\usepackage{booktabs}	
\usepackage[normalem]{ulem}
	\usepackage{enumerate}
	\usepackage{bbm}
	\usepackage{mathtools}
	\usepackage{subfig}
	\usepackage{booktabs}	
	\usepackage{setspace}
	\usepackage{graphicx}
	\usepackage{fullpage}
	
	\usepackage{soul}
	\usepackage{xspace}
	\usepackage{csquotes}
	\usepackage[margin = 1in]{geometry}
	\usepackage{placeins}
	\usepackage[footfont=normalsize,font=normalsize]{floatrow}
	\usepackage{lscape}
	\usepackage{longtable}
	\usepackage{mathabx}
    \usepackage{accents}
    \usepackage{float}

    \usepackage{tikz}
    \usetikzlibrary{calc,intersections}

	\usepackage[longnamesfirst]{natbib}
	\usepackage{bibunits}
	\defaultbibliography{Bibliography.bib}  
	\defaultbibliographystyle{aer}

	\usepackage{hyperref}
	\hypersetup{
    colorlinks=true,
    linkcolor=Red,
    filecolor=magenta,      
    urlcolor=blue,
    citecolor = Blue
}
	\usepackage{cleveref}

    \usepackage{rotating}
	\usepackage[framemethod=tikz]{mdframed}

    \usepackage{tcolorbox}
    \newtcbox{\feedback}{nobeforeafter,colframe=black,colback=white,boxrule=0.5pt,arc=2pt,
      boxsep=0pt,left=2pt,right=2pt,top=2pt,bottom=2pt,tcbox raise base}
      
    \usepackage{amsthm}

    \theoremstyle{definition}

\usepackage{array}
\newcolumntype{L}[1]{>{\raggedright\let\newline\\\arraybackslash}m{#1}}
\newcolumntype{C}[1]{>{\centering\let\newline\\\arraybackslash\hspace{0pt}}m{#1}}
\newcolumntype{R}[1]{>{\raggedleft\let\newline\\\arraybackslash\hspace{0pt}}m{#1}}

\usepackage{ragged2e}
\newlength\ubwidth

\newcommand\numberthis{\addtocounter{equation}{1}\tag{\theequation}}

\newcommand{\ubar}[1]{\underaccent{\bar}{#1}}

\title{Interpreting Event-Studies from Recent Difference-in-Differences Methods}

\author{Jonathan Roth\thanks{Brown University. \href{mailto:jonathanroth@brown.edu}{jonathanroth@brown.edu}. I am grateful to Kirill Borusyak, Clément de Chaisemartin, Xavier D'Haultf{\oe}uille, Peter Hull, Ziyi Liu, Pedro Sant'Anna, and Jesse Shapiro for comments. I thank the authors of all of the statistical packages studied here for their valuable (and often thankless) role in developing tools for implementing recent difference-in-differences methods and for making their code publicly available.}}

\begin{document}
\maketitle
\begin{abstract}
    This note discusses the interpretation of event-study plots produced by recent difference-in-differences methods. I show that even when specialized to the case of non-staggered treatment timing, the default plots produced by software for several of the most popular recent methods do not match those of traditional two-way fixed effects (TWFE) event-studies. The plots produced by the new methods may show a kink or jump at the time of treatment even when the TWFE event-study shows a straight line. This difference stems from the fact that the new methods construct the pre-treatment coefficients asymmetrically from the post-treatment coefficients. As a result, visual heuristics for evaluating violations of parallel trends using TWFE event-study plots should not be immediately applied to those from these methods. I conclude with practical recommendations for constructing and interpreting event-study plots when using these methods. 
\end{abstract}

\section{Introduction}

Modern difference-in-differences (DiD) analyses typically report an event-study plot that shows the differences in trends in the outcome between the treated and comparison groups both before and after the treatment date.\footnote{\citet{roth_pre-test_2021} identifies 70 recent papers in top economics journals displaying such plots.} Event-study plots serve two purposes. First, if the DiD design is valid, they allow the reader to evaluate the dynamic patterns of treatment effects following the implementation of the policy of interest. Second, event-study plots provide evidence about the validity of the research design. They enable the reader to perform ``visual inference'' to determine to what extent the estimated treatment effects may be driven by violations of the parallel trends assumption. An event-study plot is generally considered more convincing evidence in favor of a non-zero treatment effect if there is a discontinuity or kink in the coefficients close to the treatment date, so that the pattern seen cannot simply be explained by a smooth violation of parallel trends \citep{freyaldenhoven_visualization_2021,rambachan_more_2023}.  

Until recently, event-study plots were typically created by estimating a dynamic two-way fixed effects (TWFE) regression specification, with fixed effects for unit and time and indicators for time relative to treatment (with the coefficient for the period directly before treatment typically normalized to zero). A recent literature surveyed in \citet{de_chaisemartin_two-way-survey_2021} and \citet{roth_whats_2023} has illustrated that such TWFE specifications may be difficult to interpret in settings with heterogeneous treatment effects and staggered treatment timing, even if a parallel trends assumption holds. Most pertinently, \citet{sun_estimating_2020} showed that the event-study coefficient for relative time $r$ from a dynamic TWFE specification may be ``contaminated'' by treatment effects at a different relative time $r'$. This undermines the first purpose of event-study plots described above, since the post-treatment event-study coefficients may not correspond directly to the dynamics of the treatment effects. These issues with TWFE specifications have motivated the development of a variety of new DiD methods that allow for a more straightforward evaluation of dynamic treatment effects, assuming parallel trends holds, in settings with staggered timing and heterogeneous treatment effects. These new methods typically produce event-studies that also show pre-treatment coefficients. It would then seem natural to also use these plots for the second purpose of evaluating the extent to which the results could be explained by violations of parallel trends, using the same heuristics as with traditional dynamic TWFE specifications in the non-staggered setting. 

The main point of this note is that for the purposes of assessing violations of parallel trends, the default event-study plots produced by software for several of the most popular recent methods should not be interpreted in the same way as traditional dynamic TWFE specifications. Indeed, I show that these event-study plots do not match those of traditional dynamic TWFE specifications even when specialized to the setting with \emph{non-staggered} treatment timing. This can be seen in Figure \ref{fig:comparison} below, which shows the event-studies produced by different methods when applied to the same simulated dataset in which there is non-staggered treatment timing. In the simulation (the details of which are described below) there is no effect of the treatment but parallel trends is violated: the outcomes for the treated group are on an upwards linear trend relative to the comparison group. As expected, the event-study plot produced using dynamic TWFE yields coefficients that lie close to a straight line. By contrast, the event-study plot produced by the R package for implementing the method of \citet[][henceforth, CS]{callaway_difference--differences_2020} shows a clear kink at the treatment date.\footnote{As I describe below, the identical pattern also arises using a previous version of the R package for implementing the method of \citet[][henceforth, dCDH]{de_chaisemartin_two-way_2020}. Subsequent to the release of the initial draft of this paper in January 2024, the R package for dCDH was updated to produce results matching the TWFE event-study.} Meanwhile, the event-study plot produced by the Stata package for \citet[henceforth, BJS]{borusyak_revisiting_2024} shows a clear discontinuity at the treatment date. Given these starkly different results when applied to the \emph{same} data-set with non-staggered treatment timing, it is clear that the typical heuristics for visual inference developed based on dynamic TWFE specifications will be misleading when applied to these new estimators. 

The remainder of this note is structured as follows. In Section \ref{sec:simulation}, I provide a simple simulation illustrating how recent methods can produce very different event-study plots on the same data without staggered treatment timing. In Section \ref{sec: math}, I provide a mathematical explanation showing that these differences arise because these estimators construct the pre-treatment and post-treatment event-study coefficients asymmetrically. Section \ref{sec:recs} concludes with some recommendations and further discussion.  

\begin{figure}[!htb]
    \centering
    \subfloat[Dynamic TWFE]{\includegraphics[width = 0.5\linewidth]{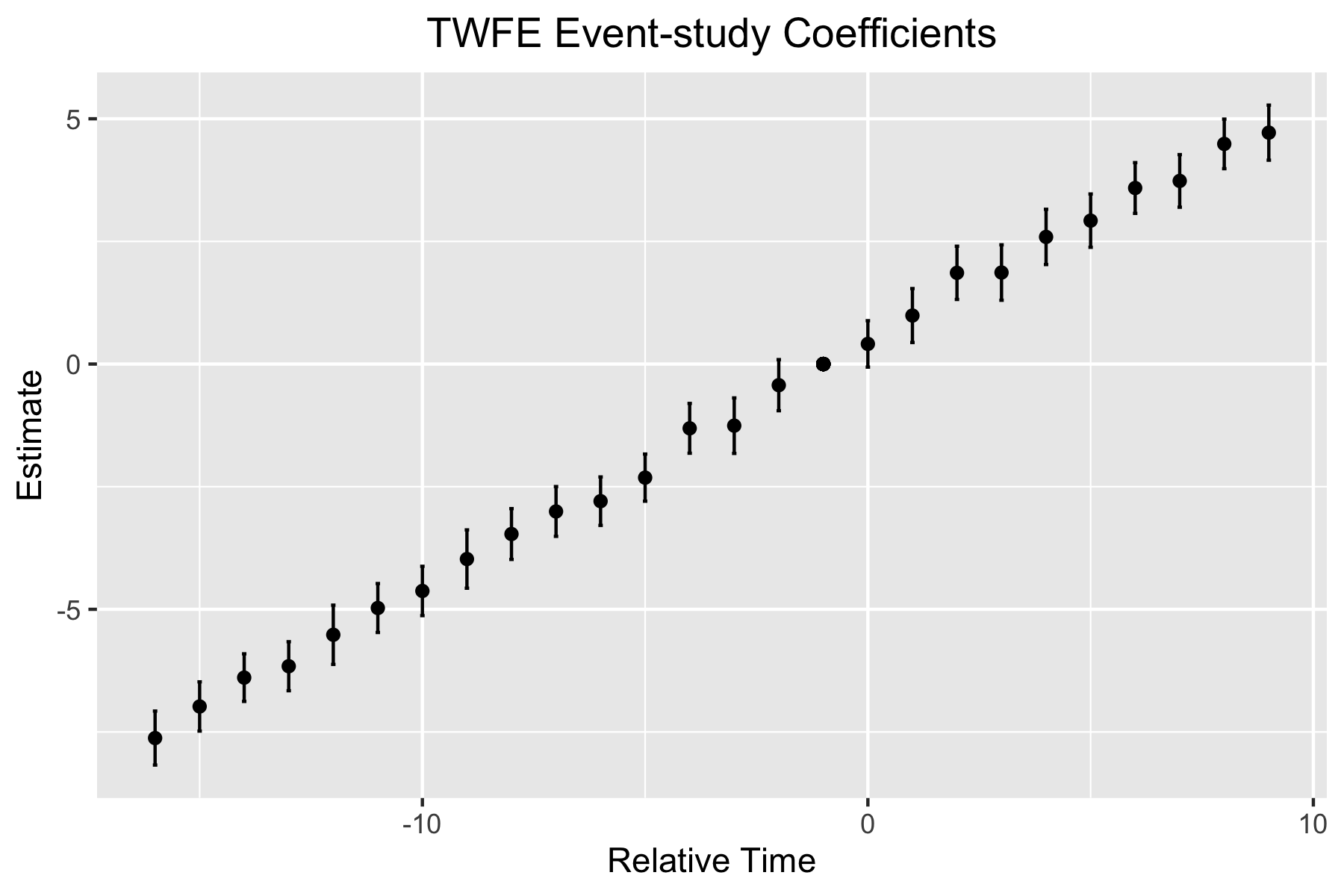}} \\
    \subfloat[CS (R package)]{\includegraphics[width = 0.5\linewidth]{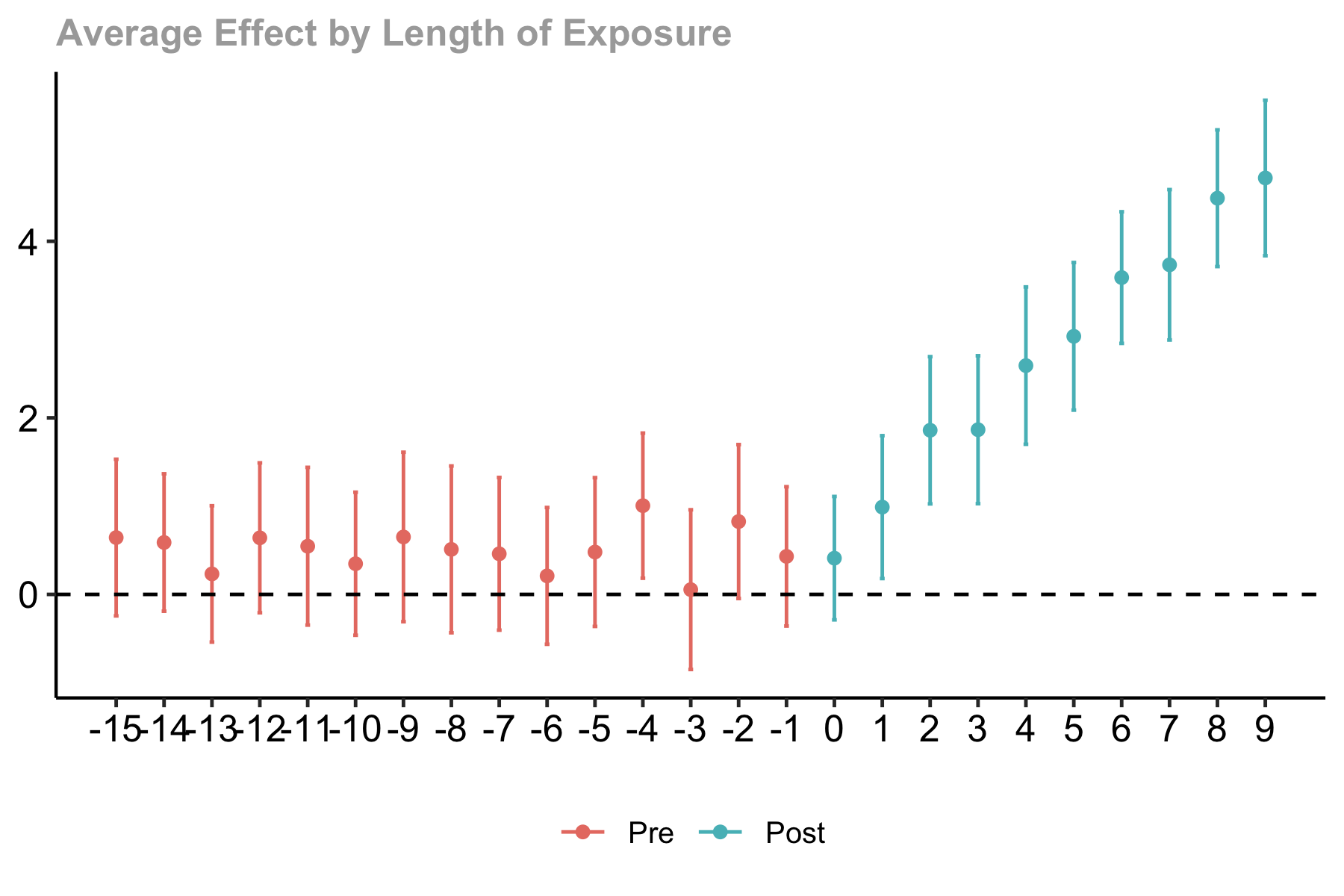}}\\
     \subfloat[BJS (Stata package)]{\includegraphics[width = 0.55\linewidth]{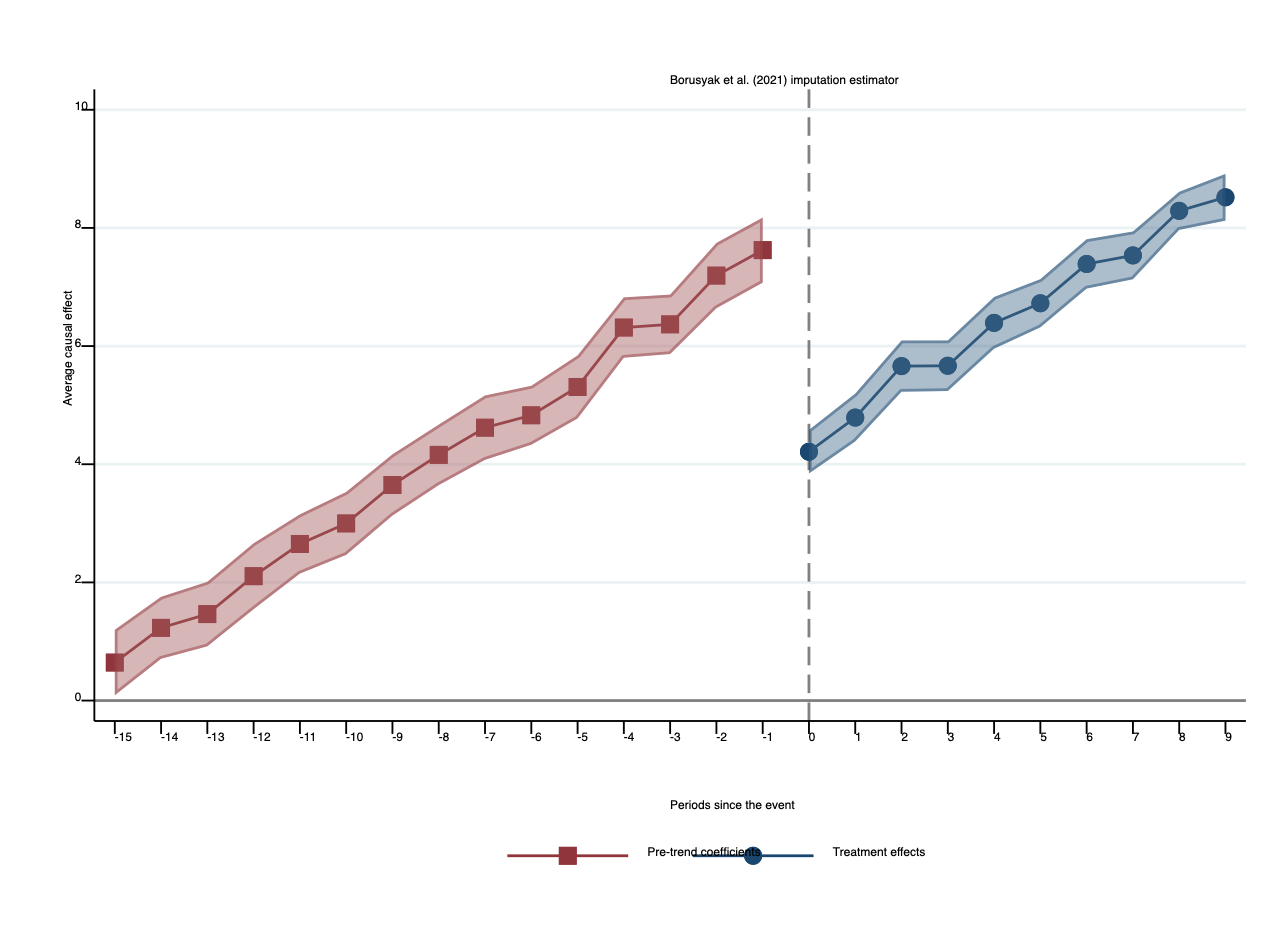}}
    \caption{Comparison of event-study plots in a non-staggered setting}
    \label{fig:comparison}
\end{figure}

\section{A simulation comparing the event-studies in a non-staggered setting\label{sec:simulation}}

I begin by illustrating the differences between the event-studies produced by TWFE and some newer methods with a simulation of an extremely simple non-staggered DiD setting.\footnote{Code for the results shown in this section is available on Github, \url{https://github.com/jonathandroth/HetEventStudies/tree/master}.} Units are either treated beginning at period $t=1$ or never-treated. We observe outcomes from period $t=-15$ to period $t=10$. There are no treatment effects in any period, so that $Y_{it}(1) = Y_{it}(0) = Y_{it}$. However, parallel trends is violated: the treated group's outcome increases linearly relative to the control group's outcome in all periods. In particular, $Y_{it}(0) = \gamma \cdot t + \epsilon_{it}$ for units in the treated group and $Y_{it}(0) = \epsilon_{it}$ for units in the control group, where $\epsilon_{it}$ are $iid$ standard normal draws. Thus, the treated group's outcome grows by $\gamma$ more than the control group's outcome on average in each period. For the illustrative simulation, I set $\gamma = 0.5$.

\begin{figure}[!hbt]
    \centering
       \subfloat[dCDH (Previous R implementation)]{\includegraphics[width = 0.5\linewidth]{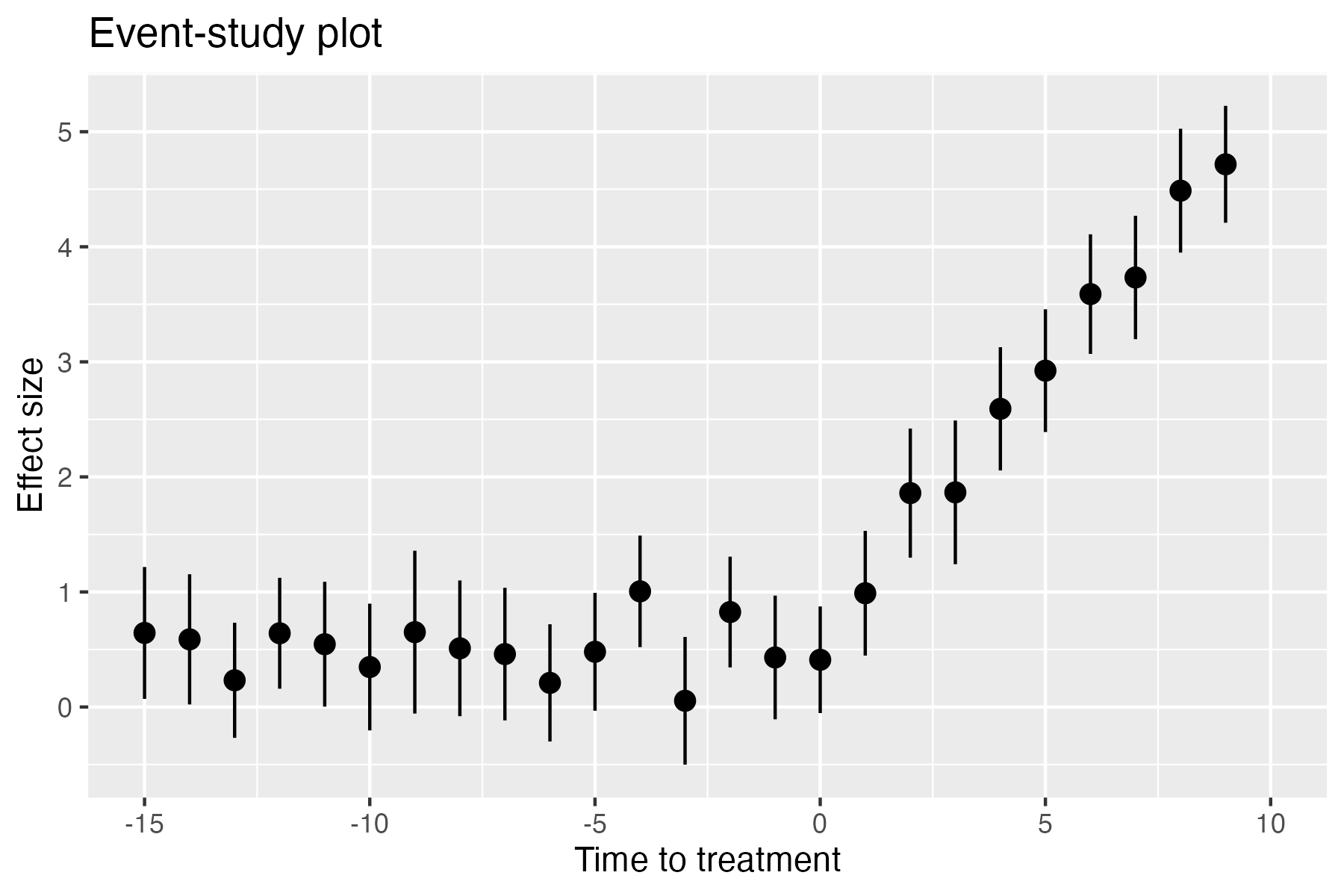}} \subfloat[dCDH (Current R implementation)]{\includegraphics[width = 0.5\linewidth]{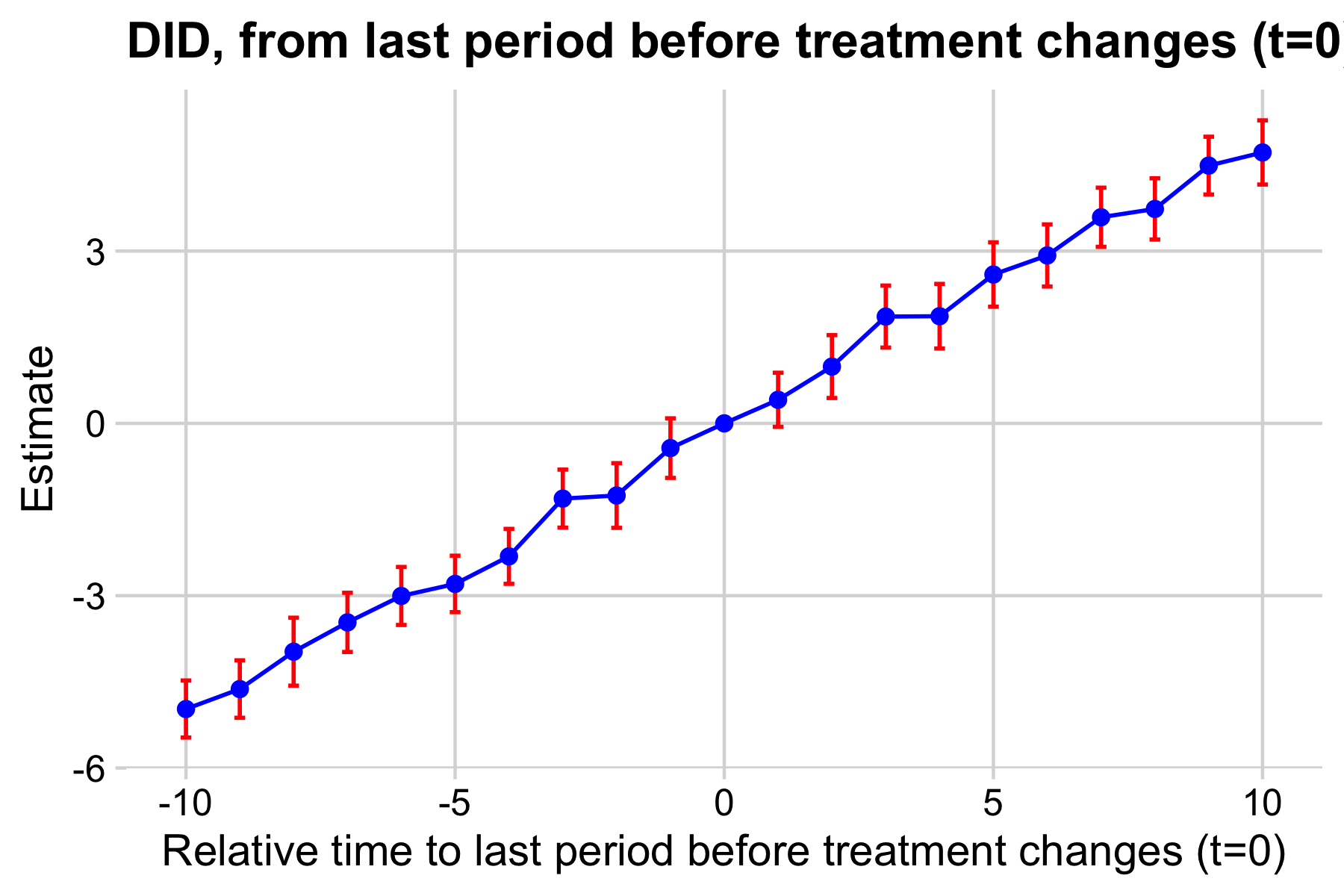}} \\
       \subfloat[Wooldridge (\texttt{etwfe} package)]{\includegraphics[width = 0.5\linewidth]{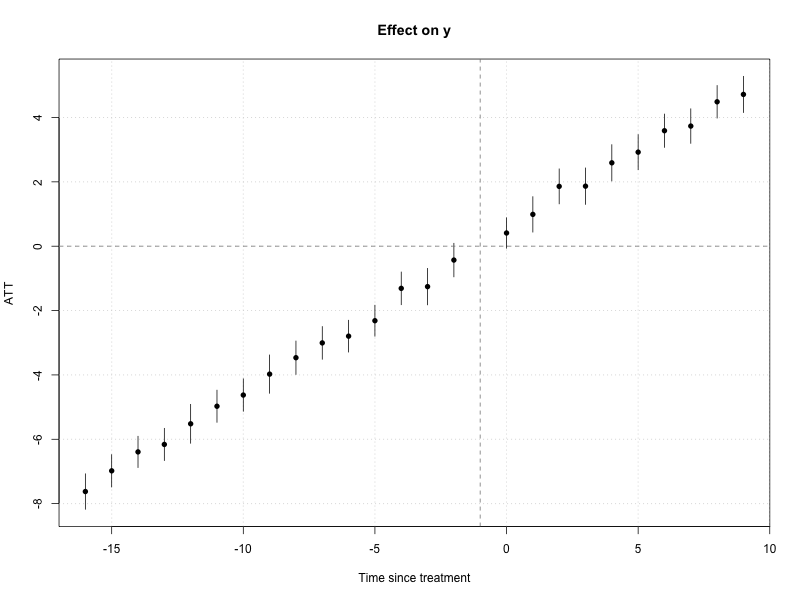}} \subfloat[SA (\texttt{fixest} implementation)]{\includegraphics[width = 0.5\linewidth]{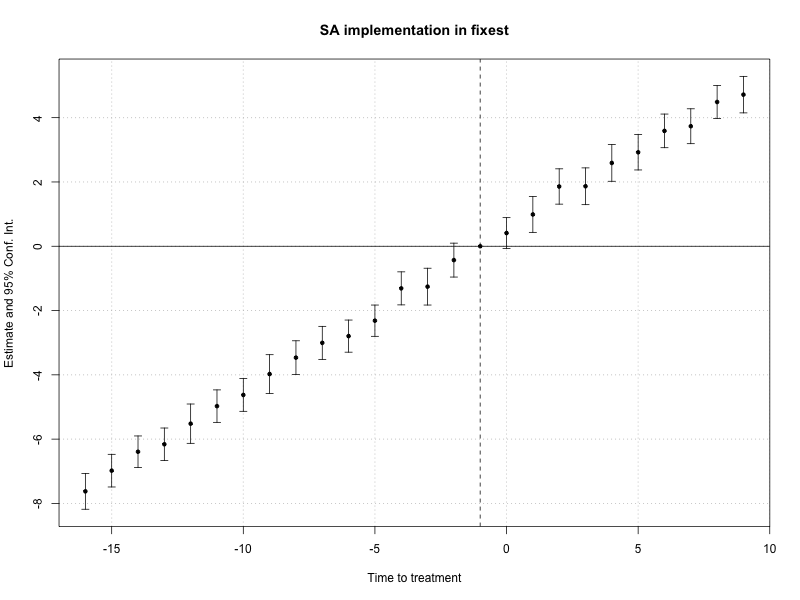}}
    \caption{Comparison of event-studies from other methods in the non-staggered setting}
    \label{fig:other-methods}
\end{figure}

Figure \ref{fig:comparison} shows event-studies for a single simulation draw from this DGP when there are 100 total units and half of them are treated. Panel (a) shows the results of a traditional dynamic TWFE ordinary least-squares specification of the form

\[Y_{it} = \alpha_i + \lambda_t + \sum_{r \neq -1} \beta_r \times 1[D_i = 1] \times 1[t=r+1] + e_{it}, \numberthis \label{eqn:twfe} \]
\noindent where $D_i$ is an indicator for treatment and $\alpha_i$ and $\lambda_t$ are unit and period fixed effects. Panel (b) shows the results for the event-study generated by the CS method, using the default settings in the \texttt{did} package in R. Panel (c) shows the event-study plot generated by the BJS method, using the default settings in the \texttt{did\_imputation} Stata package.\footnote{All packages were re-installed using the latest versions as of October 31, 2025. I obtained similar results as of the initial draft of this paper in January 2024, with one exception owing to a software update described below (see Figure \ref{fig:other-methods} and surrounding discussion).}

As described in the introduction, despite using identical data, the three methods produce very different event-study plots. The dynamic TWFE specification shows an approximately linear pre-trend that continues into the post-treatment period, with no apparent break in trend at the treatment date. By contrast, the CS event-study plot has relatively flat pre-treatment coefficients and then exhibits a kink at the time of the treatment, sloping upwards afterwards. The BJS event-study appears to show an upward pre-trend, but then exhibits a sharp downward jump at the treatment date. 

\paragraph{Other estimators.} Although I focus on CS and BJS, two of the most popular recent methods, I now comment briefly on the event-study plots produced by other recent DiD methods. At the time of the release of the initial draft of this paper in January 2024, the \texttt{DIDmultiplegt R} package for implementing the method of \citet[][dCDH]{de_chaisemartin_two-way_2020} produced identical point estimates to the CS event-study described above; see Figure \ref{fig:other-methods}. However, the most recent version of the package produces results that match the original TWFE event-study plot in my simple example.\footnote{The dCDH package only reports as many pre-trends coefficients as there are post-treatment periods. Thus, it matches the TWFE event-study for the 10 latest pre-treatment periods but does not show the earlier pre-trends coefficients.} Two other popular approaches are those of \citet[][henceforth SA]{sun_estimating_2020} and \citet[][henceforth W]{wooldridge_two-way_2025}. Both of these papers propose richer TWFE specifications to capture treatment effect heterogeneity in staggered settings. However, these TWFE specifications reduce to the usual dynamic TWFE event-study in \eqref{eqn:twfe} when specialized to the case with non-staggered treatment timing, and thus match panel (a) of Figure \ref{fig:comparison} by construction. This can be seen in panels (c) and (d) of Figure \ref{fig:other-methods}, which shows implementations of SA and W from the \texttt{etwfe} and \texttt{fixest} packages in R.\footnote{For \texttt{etwfe}, we must specify the option \texttt{cgroup = ``never''} in order to get pre-trends estimates.} Two other popular approaches are those of \citet{liu_practical_2024} and \citet{gardner_two-stage_2021}. Like BJS, these papers rely on an ``imputation'' approach for inferring counterfactual outcomes, although they construct the pre-treatment coefficients in the event-study differently. In Appendix Section \ref{appendix:insample-imp}, I discuss the results produced by the packages for these papers (\texttt{fect} and \texttt{did2s}). Both of these methods also yield an event-study that shows a jump at the date of treatment.

\section{Mathematical Examination\label{sec: math}}
In this section, I show that the differences between the event-plots created by dynamic TWFE and CS and BJS stems from the asymmetric way that the newer estimators construct the pre-treatment and post-treatment coefficients. 

I consider a slightly generalized version of the set-up in the simulation section, in which we observe panel data for periods $t = -\ubar{T},...,\bar{T}$. Treated units denoted by $D_i=1$ begin treatment at period 1, whereas comparison units ($D_i =0$) are untreated in all periods. The observed outcome is $Y_{it} = D_i Y_{it}(1) + (1-D_i) Y_{it}(0)$. For some of our results, we will consider the setting in the simulation where $Y_{it}(1) = Y_{it}(0) = \gamma \cdot t \cdot D_i + \epsilon_{it}$ and $E[\epsilon_{it} \mid D_i] = 0$. 

\subsection{Dynamic TWFE}
Consider the dynamic TWFE specification
$$Y_{it} = \alpha_i + \lambda_t + \sum_{r \neq -1} \beta_r \times 1[D_i = 1] \times 1[t=r+1] + e_{it}.$$

\noindent It is straightforward to show that the event-study coefficients estimated by OLS are given by 
$$\hat\beta_r = \left(\hat{E}[Y_{i,r+1} \mid D_i=1] -\hat{E}[Y_{i,r+1} \mid D_i=0]\right) -\left(\hat{E}[Y_{i0} \mid D_i=1] -\hat{E}[Y_{i0} \mid D_i=0]\right) ,$$
\noindent where $\hat{E}$ denotes the sample average. That is, $\hat\beta_r$ shows a 2-group, 2-period DiD estimate comparing the $D_i=1$ and $D_i=0$ groups between periods $r+1$ and $0$. Note that the formula for $\hat\beta_r$ is symmetric between pre-treatment and post-treatment periods---i.e. the same regardless of whether $r$ is greater than or less than zero.

Under mild regularity conditions, the sample means converge in probability to population means, and thus $\hat\beta_r$ is consistent for 
$$\beta_r = \left(E[Y_{i,r+1} \mid D_i=1] -E[Y_{i,r+1} \mid D_i=0]\right) -\left(E[Y_{i0} \mid D_i=1] -E[Y_{i0} \mid D_i=0]\right) .$$

\noindent Under the DGP in our simulations, we thus have

$$\beta_r = \gamma \cdot (r+1) ,$$

\noindent and so the population event-study coefficients lie on a straight line. 

\subsection{CS}
I now turn to the construction of the CS event-study estimates. CS constructs event-studies by aggregating DiD comparisons of treated and not-yet-treated units. However, by default, the construction is asymmetric across the pre-treatment and post-treatment periods. In particular, the pre-treatment coefficients are ``short-differences'', i.e. comparisons of consecutive periods, whereas the post-treatment coefficients are ``long-differences'', i.e. comparisons relative to the period before treatment. 

Formally, in our setting with common treatment timing, we have that 
\begin{equation*}
\hat\beta_r^{CS} = \begin{cases} \left(\hat{E}[Y_{i,r+1} \mid D_i = 1]  - \hat{E}[Y_{i,r+1} \mid D_i = 0]\right) - \left(\hat{E}[Y_{i,r} \mid D_i = 1]  - \hat{E}[Y_{i,r} \mid D_i = 0]\right)   &\text{ if } r <0 \\
 \left(\hat{E}[Y_{i,r+1} \mid D_i = 1]  - \hat{E}[Y_{i,r+1} \mid D_i = 0]\right) - \left(\hat{E}[Y_{i0} \mid D_i = 1]  - \hat{E}[Y_{i0} \mid D_i = 0]\right)  &\text{ if } r \geq 0. \end{cases}
\end{equation*}

\noindent Note that for pre-treatment coefficients ($r < 0$), we compare outcomes in consecutive periods (periods $r+1$ vs $r$), whereas for post-treatment coefficients ($r \geq 0$), we compare outcomes over longer horizons (periods $r+1$ vs $0$). Under mild regularity conditions, it follows that $\hat\beta_r^{CS}$ is consistent for 
\begin{equation*}
\beta_r^{CS} = \begin{cases} \left(E[Y_{i,r+1} \mid D_i = 1]  - E[Y_{i,r+1} \mid D_i = 0]\right) - \left(E[Y_{i,r} \mid D_i = 1]  - E[Y_{i,r} \mid D_i = 0]\right)   &\text{ if } r <0 \\
 \left(E[Y_{i,r+1} \mid D_i = 1]  - E[Y_{i,r+1} \mid D_i = 0]\right) - \left(E[Y_{i0} \mid D_i = 1]  - E[Y_{i0} \mid D_i = 0]\right)  &\text{ if } r \geq 0 .\end{cases}
\end{equation*}
\noindent In our simulation DGP, we thus have that 

\begin{equation*}
\beta_r^{CS} = \begin{cases} \gamma  &\text{ if } r <0 \\
 \gamma \cdot (r+1)  &\text{ if } r \geq 0 \end{cases}
\end{equation*}

\noindent so the population version of the CS event-study coefficients show a kink at zero if $\gamma \neq 0$. 

\subsection{BJS}
Like the CS event-study approach, the BJS event-study is also constructed in a way that is asymmetric between the pre-treatment and post-treatment periods. For the post-treatment effects, they use what they call an imputation approach: they fit a TWFE regression using only untreated $(i,t)$ pairs, then form individual treatment effect estimates of the form $\hat\tau_{it} = Y_{it} - \hat{Y}_{it}$, where $\hat{Y}_{it}$ is the prediction from the TWFE regression. They then average the $\hat\tau_{it}$ for treated units at a given lag from treatment. For the pre-treatment coefficients, they simply run a dynamic TWFE regression using only $(i,t)$ pairs that are untreated, with dynamic indicators for the number of periods until treatment (with the earliest pre-treatment period normalized to zero).\footnote{The Stata package requires the researcher to specify how many pre-treatment coefficients they would like to calculate, and all earlier periods are used as the omitted category. The omitted category is thus the earliest period if one requests $\ubar{T}$ pre-treatment coefficients.}  

In our special case of non-staggered treatment timing, the post-treatment event-study coefficients for BJS correspond to simple DiDs between period $r+1$ and the \emph{average} outcome in the pre-treatment period. By contrast, the pre-treatment coefficients correspond to DiDs between period $r+1$ and the earliest period, $-\ubar{T}$. Specifically, let $\bar{Y}_i^{pre} = \frac{1}{\ubar{T}+1} \sum_{t=-\ubar{T}}^{0} Y_{it}$ be $i$'s average outcome in periods $t \leq 0$. Then we have 
\begin{equation*}
\hat\beta_r^{BJS} = \begin{cases} \left(\hat{E}[Y_{i,r+1} \mid D_i = 1]  - \hat{E}[Y_{i,r+1} \mid D_i = 0]\right) - \left(\hat{E}[Y_{i,-\ubar{T}} \mid D_i = 1]  - \hat{E}[Y_{i,-\ubar{T}} \mid D_i = 0]\right)   &\text{ if } r <0 \\
 \left(\hat{E}[Y_{i,r+1} \mid D_i = 1]  - \hat{E}[Y_{i,r+1} \mid D_i = 0]\right) - \left(\hat{E}[\bar{Y}_{i}^{pre} \mid D_i = 1]  - \hat{E}[\bar{Y}_{i}^{pre} \mid D_i = 0]\right)  &\text{ if } r \geq 0. \end{cases}
\end{equation*}
\noindent This makes clear that the construction is asymmetric, with $\hat\beta_r^{BJS}$ using a comparison to period $-\ubar{T}$ for the pre-treatment coefficients and to the average of the pre-treatment periods for the post-treatment coefficients. 

Under mild regularity conditions, we thus have that $\hat\beta_r^{BJS}$ is consistent for
\begin{equation*}
\beta_r^{BJS} = \begin{cases} \left(E[Y_{i,r+1} \mid D_i = 1]  - E[Y_{i,r+1} \mid D_i = 0]\right) - \left(E[Y_{i,-\ubar{T}} \mid D_i = 1]  - E[Y_{i,-\ubar{T}} \mid D_i = 0]\right)   &\text{ if } r <0 \\
 \left(E[Y_{i,r+1} \mid D_i = 1]  - E[Y_{i,r+1} \mid D_i = 0]\right) - \left(E[\bar{Y}_{i}^{pre} \mid D_i = 1]  - E[\bar{Y}_{i}^{pre} \mid D_i = 0]\right)  &\text{ if } r \geq 0. \end{cases}
\end{equation*}

\noindent Under our simulation DGP, this corresponds to\footnote{In the calculation below, we use the fact that $$E[\bar{Y}_i^{pre} \mid D_i=1] = \gamma \frac{1}{\ubar{T}+1} (-\ubar{T} + ... + 0) = \gamma \frac{1}{\ubar{T} + 1} \left(- \frac{1}{2}\ubar{T} (\ubar{T}+1)\right) =-\frac{1}{2} \ubar{T} \cdot \gamma.$$}

\begin{equation*}
\beta_r^{BJS} = \begin{cases} \gamma \cdot (r+1 - (-\ubar{T}))  &\text{ if } r <0 \\
 \gamma \cdot (r+1 - \frac{1}{2}  (-\ubar{T}))  &\text{ if } r \geq 0 \end{cases}
\end{equation*}

\noindent so the population version of BJS will exhibit a jump at zero if $\gamma \neq 0$.

\section{Practical takeaways and recommendations\label{sec:recs}}
The first practical takeaway from these results is that the default event-studies from the methods considered in Figure \ref{fig:comparison} should not be interpreted in the same way as traditional dynamic TWFE event-study plots for the purposes of evaluating parallel trends violations. Owing to the asymmetric construction of the pre-treatment and post-treatment coefficients, a kink or jump in the plot may arise even if there is no treatment effect and parallel trends is equally violated in all periods. One should therefore not apply the typical heuristics for visual inference---e.g. looking for a discontinuity or kink at the treatment date---that may be familiar from TWFE event-studies. Sensitivity analyses that compare the pre-treatment and post-treatment coefficients from these event-studies \citep[e.g.][]{manski_how_2017, rambachan_more_2023} may likewise yield misleading conclusions given the asymmetric construction of the pre- and post-treatment coefficients.

A natural follow-up question is: to what extent can the event-study plots for these alternative methods be modified to be more comparable to conventional event-study plots? I discuss CS and BJS in turn, followed by some discussion of additional considerations.

\paragraph{CS.} For CS, there is a straightforward answer, which is to use ``long-differences'' for the pre-treatment coefficients as well as the post-treatment coefficients (i.e. always use the period before treatment as the baseline). This can be implemented, for example, in the \texttt{did} R and \texttt{csdid} Stata packages using the options \texttt{base\_period = ``universal''} and \texttt{long2}, respectively.\footnote{A blog post by \citet{callaway_universal_2021} provides further discussion of the differences between these two options. Similar to our ongoing example, he considers a DGP with a linear violation of parallel trends, although it only has one post-treatment period and thus does not exhibit the kink seen in our example. Nevertheless, he concludes that long-run violations of parallel trends are easier to visualize using the universal-base option, but argues that using short-differences may make it easier to detect violations of the no anticipation assumption.} Using these settings, the event-study estimates are numerically equivalent to the dynamic TWFE specification in the non-staggered setting considered here. 

\paragraph{BJS.} For BJS, the answers are somewhat less obvious. The challenge arises from the fact that, by construction, BJS uses \emph{all} of the pre-treatment data in predicting post-treatment counterfactuals for treated units (as they show, this is efficient under spherical errors).\footnote{The same is true of other imputation approaches such as \citet{gardner_two-stage_2021} and \citet{liu_practical_2024}. In our non-staggered setting, the post-treatment effects estimated by these approaches are equivalent to those for BJS; the methods only differ in how they create the pre-treatment coefficients. Thus, the recommendations for modifying the pre-treatment coefficients for BJS also apply to these other methods.} Any analysis that does a similar imputation in the pre-treatment period will either (a) be looking at \emph{in-sample} prediction fit---i.e. looking at residuals $Y_{it} - \hat{Y}_{it}$, where $Y_{it}$ was included in the training data for creating $\hat{Y}_{it}$, or (b) using predictions trained only on a sub-set of the pre-treatment data, in contrast to the predictions for post-treatment outcomes, which use all of the pre-treatment data. Nevertheless, it may be possible to construct event-study plots using imputation estimators that are more similar to the standard TWFE event-study plot than the BJS implementation. Below I discuss a few possible options, along with their pros and cons; this strikes me as an area where additional innovation in future work may also be possible. 

In my ongoing non-staggered example, one natural approach to make the BJS event-study coefficients more symmetric would be to use the \emph{average} of the pre-treatment periods as the reference for both the pre-treatment and post-treatment event-study coefficients. That is, we could construct 
\begin{equation*}
\hat\beta_r^{BJS,new} =\left(\hat{E}[Y_{i,r+1} \mid D_i = 1]  - \hat{E}[Y_{i,r+1} \mid D_i = 0]\right) - \left(\hat{E}[\bar{Y}_{i}^{pre} \mid D_i = 1]  - \hat{E}[\bar{Y}_{i}^{pre} \mid D_i = 0]\right) \text{ for all $r$}.
\end{equation*}
\noindent This would match the usual TWFE event-study up to a \emph{vertical shift} of all the coefficients. That is, $\hat{\beta}^{BJS,new}_{r} = \hat{\beta}^{TWFE}_{r} + \hat{c}$ for a value $\hat{c}$ that doesn't depend on $r$. However, it is worth noting that this construction guarantees that the modified pre-trends coefficients \emph{average} to zero (i.e. $\sum_{r < 0} \hat{\beta}^{BJS,new}_r = 0$). Visual inference based on comparing the \emph{average} of the event-study coefficients before and after treatment may thus be misleading, since the average in the pre-treatment period is zero by construction. Nonetheless, using this more symmetric imputation approach seems preferable to me than the default BJS approach if one wants output resembling a typical event-study, provided that the researcher keeps in mind that they should only be looking at the relative differences between the coefficients and not their averages before and after treatment. How can this approach be extended to the staggered setting? \citet{liu_cohort-anchored_2025} proposes a notion which he refers to as ``block bias'' which measures the pre-treatment violations of parallel trends for each adoption cohort (relative to a fixed comparison group) in a staggered setting.\footnote{\citet{liu_cohort-anchored_2025} also discusses how these block biases can be used in conjunction with the sensitivity analysis approach of \citet{rambachan_more_2023}.} Applied to our non-staggered example, the pre-trends estimated by the block bias approach align with the modified estimator $\hat\beta_r^{BJS,new}$. Perhaps the most natural approach for imputation estimators in staggered settings is then to create an event-study where the pre-trends estimates are formed by taking a (weighted) average of the block biases across adoption cohorts.

A related suggestion by \citet{li_benchmarking_2025} is to adopt a ``leave-one-out'' approach to pre-trends estimation using imputation estimators, where for relative time $r < 0$, one computes averages of the form $Y_{it} - \hat{Y}_{it}^{loo}$, where the prediction $\hat{Y}_{it}^{loo}$ does not use information from relative time $r$. This approach does not align with the standard TWFE event-study when specialized to the common timing case---the pre-trend coefficients are based on predictions using all-but-one pre-treatment period but the post-treatment coefficients are based on predictions using all the pre-treatment periods---but the differences are small when the number of pre-treatment periods is large.\footnote{Another approach, which is implemented in the \texttt{fect} and \texttt{did2s} packages, reports averages of the form $Y_{it} - \hat{Y}_{it}$ for \emph{all} relative time periods $r$, where $\hat{Y}_{it}$ is the same as for BJS. As discussed in more detail in Appendix Section \ref{appendix:insample-imp}, this produces an event-study that shrinks the coefficients $\hat{\beta}^{BJS,new}_{r}$ towards zero in the pre-treatment period, and thus may produce event-study plots with discontinuities at the treatment date.} 

An alternative approach recommended by BJS is to completely separate the testing and estimation steps.\footnote{The documentation for BJS' Stata package states, ``A pre-trend test (for the assumptions of parallel trends and no anticipation) is a separate exercise.''} The pre-treatment coefficients in the BJS default event-study are still valid tests of pre-treatment parallel trends in the sense that if parallel trends holds in all pre-treatment periods, the $\beta_r^{BJS}$ should be zero for all $r<0$, and so can be used for testing the null of parallel pre-trends.\footnote{See \citet{roth_pre-test_2021} and \citet{bilinski_seeking_2018} regarding pitfalls of simply looking at the significance of the pre-treatment coefficients, however. As argued in those papers, researchers should ideally also look at the \emph{magnitudes} of the pre-trends estimates and their confidence intervals. However, evaluating the magnitudes of potential bias from pre-trends is challenging if the pre-treatment and post-treatment coefficients are not constructed symmetrically.} However, if one adopts this approach, then I would humbly suggest putting the BJS pre-treatment estimates on a different plot from the post-treatment estimates to avoid making misleading visual inferences. If one prefers to have a typical-looking event-study plot, then a straightforward way of separating the testing and estimation steps would be for the researcher to report an event-study for a method that treats the periods completely symmetrically (e.g. using the \texttt{did} package with the options described above) to visually evaluate the plausibility of the parallel trends assumption, and to then subsequently report the potentially more-precise post-treatment estimates from BJS that assume parallel trends in all periods. One can also conduct sensitivity analysis for these post-treatment estimates directly using the approach in \citet{rambachan_more_2023}; see \citet{liu_cohort-anchored_2025} for suggestions for implementing this approach in staggered settings in conjunction with imputation estimators like BJS.

\paragraph{Additional discussion.} It is worth emphasizing that the discussion here does not threaten the validity of the post-treatment event-study estimates for recent methods \emph{if} the appropriate parallel trends assumption holds: in this case, all of the recent methods yield interpretable treatment effect estimates even under heterogeneous treatment effects. Rather, the asymmetric construction of the pre-treatment and post-treatment event-study coefficients makes it challenging to visualize whether the estimated post-treatment coefficients could be explained by a violation of parallel trends rather than a treatment effect, at least using the conventional visual heuristics (or typical sensitivity analyses).

Finally, it is worth noting that even if a method produces an event-study plot that matches the usual TWFE event-study when specialized to the non-staggered setting, additional complications may arise when interpreting these results in staggered settings. One particularly important issue for interpretation in the  staggered setting, which is discussed in detail in \citet{liu_cohort-anchored_2025}, is that the composition of units that contribute to an event-study coefficient will often change across different relative time periods. Indeed, if a panel is balanced in calendar time, then with staggered treatment adoption it is necessarily \emph{imbalanced} in relative time. One therefore has to be careful in interpreting aggregated event-studies since the composition of units that contribute to the coefficient for relative time $r$ will be different from that for relative time $r'$. To assess the importance of changing composition, it will often be useful to supplement aggregated event-studies with plots of the time series of the mean outcome by adoption cohort. It may also be useful to report aggregated event-studies that enforce some balancing criteria---e.g. reporting an event-study that is restricted to units who are observed at least 3 periods before and after treatment.

\bibliography{bibliography}

\clearpage
\appendix 

 \renewcommand{\figurename}{Appendix Figure}
\setcounter{figure}{0}

\section{Event-study plots based on in-sample imputation\label{appendix:insample-imp}}

\begin{figure}[!hb]
    \centering
    \subfloat[\texttt{fect}]{ \includegraphics[width = 0.5\linewidth]{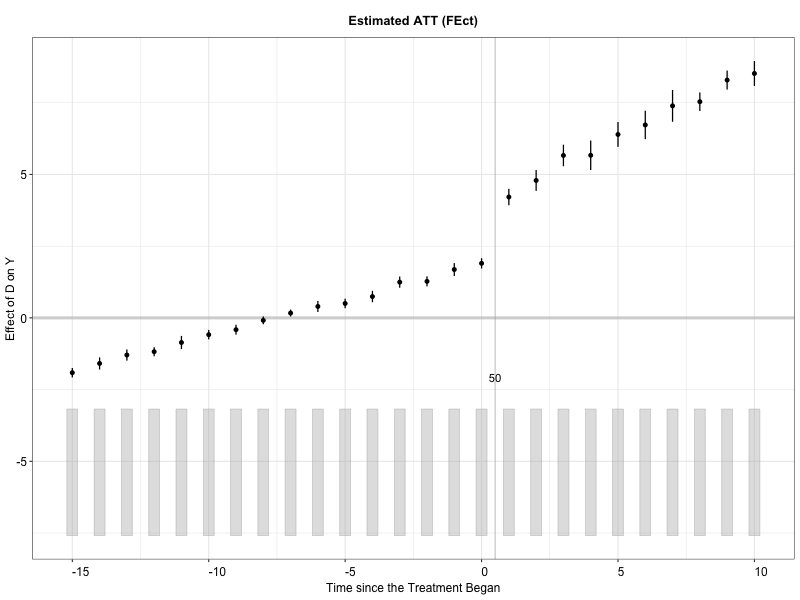} }    \subfloat[\texttt{did2s}]{ \includegraphics[width = 0.5\linewidth]{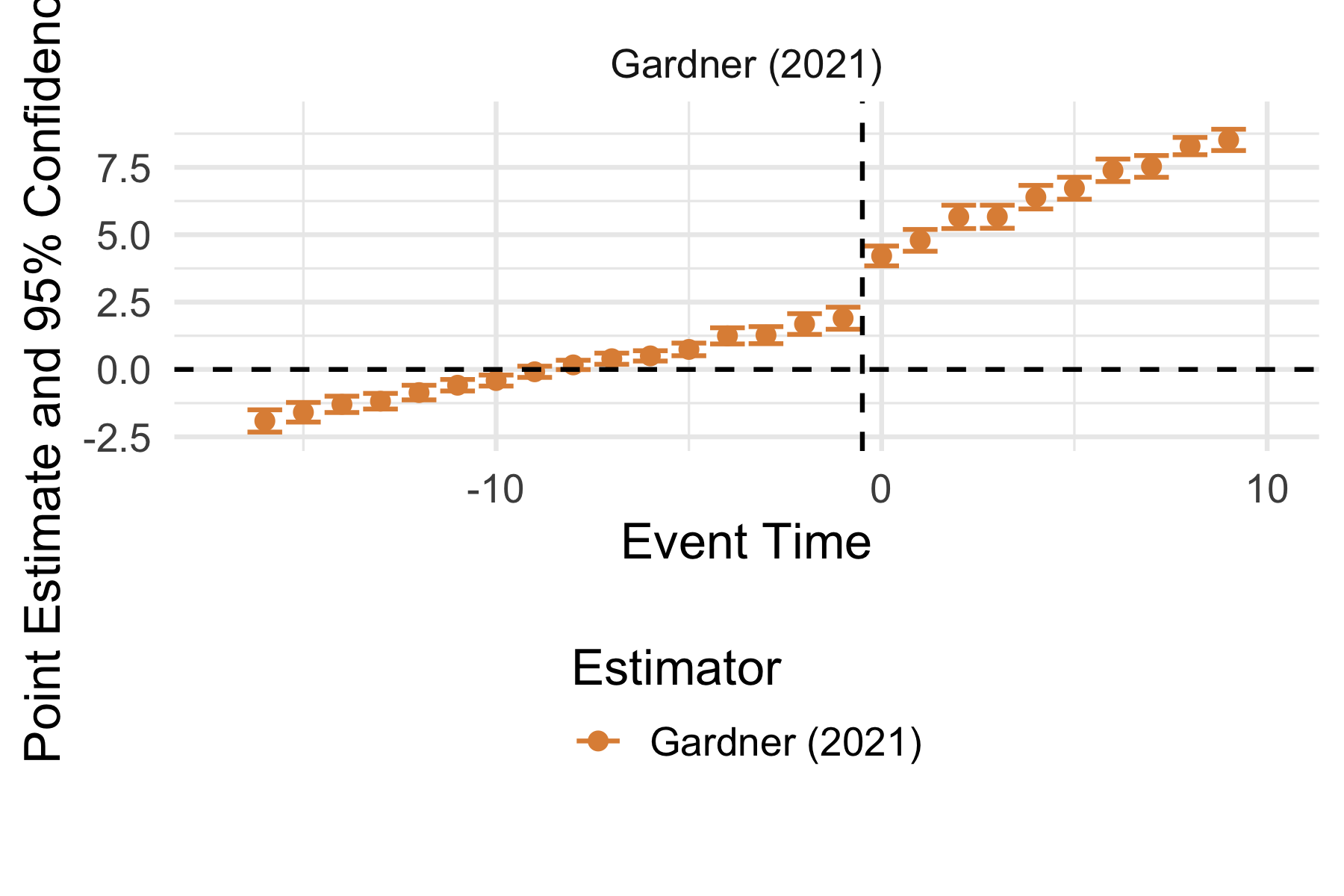} }
    \caption{Event-studies from methods using in-sample imputation (\texttt{fect} and \texttt{did2s})}
    \label{fig:insample-imp}
\end{figure}

Another approach to constructing event-studies for imputation estimators, which is implemented in the \texttt{fect} and \texttt{did2s} packages, reports averages of the form $Y_{it} - \hat{Y}_{it}$ for \emph{all} relative time periods $r$, where $\hat{Y}_{it}$ is the same as for BJS described above. \citet{li_benchmarking_2025} study this approach in detail, and argue that it systematically under-estimates the magnitude of the pre-trends. The issue stems from the fact that the pre-treatment outcomes for treated units are used to form $\hat{Y}_{it}$, and thus $Y_{it} - \hat{Y}_{it}$ is an \emph{in-sample} prediction error for pre-treatment observations. When applied to the non-staggered setting, the results in \citet{li_benchmarking_2025} imply that this in-sample approach yields pre-treatment coefficients equal to $\frac{N_0}{N} \hat\beta^{BJS,new}_r$ for $r < 0$, where $N_0$ is the number of untreated units. Thus, the pre-treatment coefficients $\hat\beta^{BJS,new}_r$ are multiplied by the factor $\frac{N_0}{N} < 1$, which is very small when most of the units are treated. Appendix Figure \ref{fig:insample-imp} illustrates this attenuation by plotting the event-studies produced by \texttt{fect} and \texttt{did2s} in our simulated data: there is a clear jump in the coefficients around the treatment date.

\nocite{fixest,did,did2s,DIDmultiplegt,etwfe,fect,did_imputation}

\end{document}